# Pinhole induced efficiency variation in perovskite solar cells

Sumanshu Agarwal[1] and Pradeep R. Nair[2]

[1]Department of Energy Science and Engineering, [2]Department of Electrical Engineering

Indian Institute of Technology Bombay, Mumbai, Maharashtra, 400076, India

sumanshu@iitb.ac.in, prnair@ee.iitb.ac.in

*Abstract*:

Process induced efficiency variation is a major concern for all thin film solar cells, including the emerging perovskite based solar cells. In this manuscript, we address the effect of pinholes or process induced surface coverage aspects on the efficiency of such solar cells through detailed numerical simulations. Interestingly, we find the pinhole size distribution affects the short circuit current and open circuit voltage in contrasting manners. Specifically, while the $J_{SC}$ is heavily dependent on the pinhole size distribution, surprisingly, the $V_{OC}$ seems to be only nominally affected by it. Further, our simulations also indicate that, with appropriate interface engineering, it is indeed possible to design a nanostructured device with efficiencies comparable to that of ideal planar structures. Additionally, we propose a simple technique based on terminal I –V characteristics to estimate the surface coverage in perovskite solar cells.

# 1. Introduction

Perovskite solar cells have gained immense research interest in last few years mainly because of its high energy conversion efficiency[1,2]. Though state of the art efficiency for this class of solar cells is more than 20%[3,4], large variations in performances of the devices fabricated in different laboratories have been reported[5–13]. Poor morphological control[14,15] and bad surface coverage of perovskite between electron and hole transport layers (ETL and HTL)[1,16] are known to have adverse effects on the performance of the devices (see Fig. 1). There have been several attempts to deposit almost pinhole free and smooth perovskite films using techniques like co-evaporation[1], use of PbAc$_2$[17,18], PVP (poly-vinylpyrrolidone) as surfactant[19], variation in the anneal temperature[15,20] and anneal time[21], but quantitative estimates of losses due to sub-optimal surface coverage are still not available in literature.

In this manuscript we discuss the effect of pinholes or non-ideal perovskite surface coverage on the efficiency through detailed optical and carrier transport simulations which are supported through an analytical model as well. While degradation in the performance metrics is generally anticipated due to poor surface coverage or pinholes, here we show that, surprisingly, the $V_{OC}$ is rather independent of the pinhole size distribution and is more affected by the net surface coverage. However, the $J_{SC}$ of the device is indeed affected by size distribution of the pinholes as well as by net surface coverage. Further, carrier recombination at ETL/HTL interface (due to the absence of perovskite between them) can have interesting implications, including near ideal performance for the devices with sub-optimal surface coverage. Below, we first describe the model system to study the effect of surface coverage on the performance of device. The model is then extended to explore the effects of bad interface between perovskite and HTL. We also propose simple schemes to estimate surface coverage from terminal I-V characteristics.

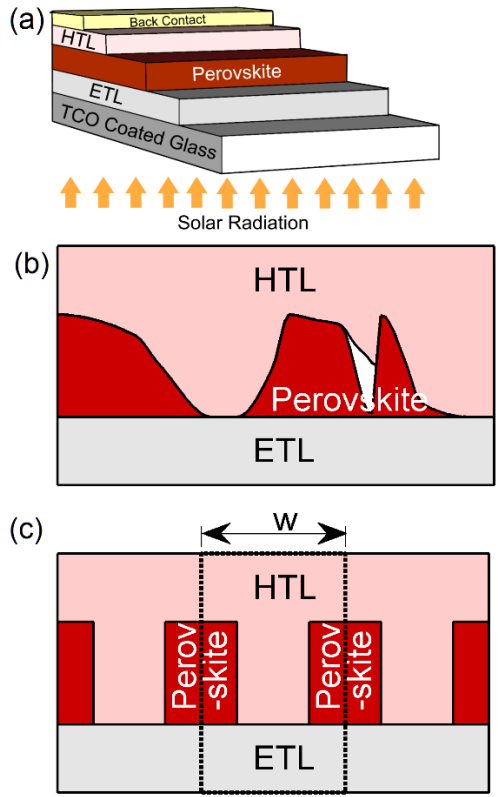

Fig 1: Schematic of perovskite solar cell. (a) The layer by layer structure of a typical perovskite solar cell. (b) A schematic representation of non-ideal surface coverage in perovskite solar cells. (c) The device structure used for numerical simulations. '**w**' is the width of unit cell which indicated by dashed line.

## 2. Model System

The structure of a perovskite solar cell is shown in figure 1 (a). Here, the perovskite is sandwiched between ETL and HTL. Non-uniform deposition of perovskite between ETL and HTL leads to empty space or direct contact between ETL and HTL as shown in cross-sectional view of perovskite solar cell (fig. 1b). These regions could result in increased recombination and affect the photon absorption properties as well. The fraction of area covered by perovskite between ETL/HTL is denoted as surface coverage (s) and the region not covered by perovskite is referred as void in this paper. Typically, poor surface coverage of perovskite between ETL/HTL causes loss in performance due to: a) less absorption of photons and b) increase in the recombination at the interface[15].

As mentioned before, current literature lacks quantitative estimates of such surface coverage aspects which are required to accurately model the eventual device performance. Given this scenario, here we make a few simplifying assumptions such that the problem becomes computationally and conceptually tractable – (a) all voids are uniform in size and (b) all voids are filled with HTL. The corresponding scheme is shown in part (c) of the figure 1, which is a periodic structure whose unit cell is indicated by the dashed rectangle. Accordingly, any real device can be then considered as combination of such unit cells (horizontally) of many sizes and surface coverage. A unit cell can be uniquely defined using the variables: a) width of the unit cell ($w$) and b) surface coverage ($s$) (see fig. 1c). As per this definition, the size of a void is given by the term $w \times s$.

We calculate optical and electrical characteristics of the device defined by such unit cells to explore the effects of non-uniform surface coverage. Optical characteristics of device is obtained through numerical solution of Maxwell's equations in frequency domain[22] (eq. 1)

$$\nabla \times (\nabla \times E_{em}) = \kappa_0^2 (n_r - ik)^2 E_{em}, \qquad (1)$$

where $E_{em}$ is the electric field of electromagnetic (EM) wave, $\kappa_0$ is the wave number of the free space, $n_r$ and $k$ are real and imaginary parts of refractive index, respectively. $\kappa_0$ is defined as $\kappa_0 = \omega/c_0$, where $\omega$ is angular velocity of light and $c_0$ is the speed of light in free space. EM power at any point is $P = \frac{1}{2}\epsilon_0 c |E_{em}|^2$. The absorptance for a particular wavelength is given by the ratio of power dissipated in the perovskite to the incident power for that wavelength. Here we assume that each absorbed photon results in the generation of one free e-h pair in perovskite.

The electrical (JV) characteristics under dark and illuminated conditions are simulated through self-consistent solutions of drift-diffusion and Poisson's equations (see supplementary material for details). We have assumed $TiO_2$ as ETL (225 nm thick) and spiro-MeOTAD as HTL (200 nm thick) besides 300 nm thick perovskite. The material parameters are adopted from literature[23–25] and calibrated using detailed simulations, reported elsewhere.[26] For the electrical calculations we use uniform photo-generation rate inside the perovskite which is equivalent to generation rate obtained from optical simulations. We consider SRH, radiative, and Auger recombination mechanisms in perovskite along with trap assisted recombination at ETL/HTL interface, while the bulk ETL and HTL are treated as recombination/generation free. All our simulations are performed using doped contact layers unless otherwise mentioned.

As the propagation of light depends on void size (i.e, $w \times s$), both $w$ and $s$ are of critical importance for optical calculations. For optical simulations we have varied unit cell width ($w$) from 100 nm to 10 μm and surface coverage ($s$) is varied from 0% to 100%. Unlike the optics, the carrier transport is critically influenced by the electric field profiles and the carrier mobility. Our simulations indicate the absence of any significant electric field fringing around the voids as the perovskite and HTL dielectric constants are assumed to be of the same order. Moreover, the carrier collection lengths are typically much larger than the active layer thickness. Hence the JV characteristics depend primarily on $s$ rather $w$. So we have assumed '$w$' as 1 μm for electrical transport simulations. These assumptions are supported through numerical simulations.

## 3 Results and discussions

### 3.1 Optical Characteristics:

Numerical solutions[27] of eq. 1 for 100% surface coverage show that only about 80% of solar energy of interest (*i.e.*, more than perovskite band gap energy) is absorbed by perovskite while rest is lost either by reflection or parasitic absorption (see fig S1 for details). These results are in close agreement with previous reports based on transfer matrix method.[28] For structures with $s < 100\%$, diffraction of light becomes an important phenomena that could influence the carrier generation rate in perovskites (see fig. S2). Figure 2a compares the volumetric integration of photons absorbed in the perovskite vs wavelength for different surface coverages (w = 1 μm). The values are normalized with the results of device with $s = 100\%$. Here we observe that photons absorbed in the perovskite increases with surface coverage, an obvious result. We also observe that this increase is nonlinear with 's' for moderate to large wavelength regime (more than 450 nm). Further we find that diffraction pattern is different for different wavelengths and surface coverage (see fig 2 and supplementary material fig S2 and S3).

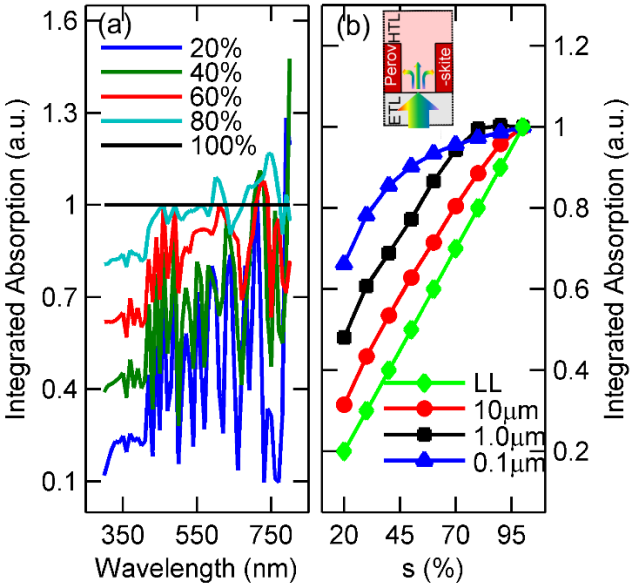

Fig 2: Effect of surface coverage on the optical characteristics of the device. (a) The integrated absorption of photons inside the perovskite ($w = 1$ μm) vs wavelength for different surface coverages while the integrated absorption rates of solar spectrum (300-800 nm) inside the perovskite vs surface coverage for different width of unit cells are shown in part (b).

Figure 2b shows the effect of unit cell width on the net absorption of photons in perovskite under 1 sun illumination (normalized w.r.t. device with $s = 100\%$). Here, we observe that– a) absorbed photons, in general, increases with reduction in unit cell width, b) for the large width of unit cell, total photons absorbed in the active layer varies linearly with surface coverage (results for w = 10 μm), and c) lower limit of absorbed photons is linear with 's' (shown by diamonds, see supplementary material for detailed calculations). Note that the lower limit for the generation rate (or equivalently the absorption rate) ($G$) in the perovskite is given by the relation $G(s) = s \times G(100)$ where $G(100)$ is the generation rate of e-h pairs (s$^{-1}$) in perovskite for a device with 100% surface coverage. This limit assumes that negligible fraction of photons incident on the void are diffracted towards the active layer.

All the above observations follow the fact that the ratio of the light diffracted to the total light incident on the void decreases with increase in void size (void size = w × s, also see inset in fig 2b). It is expected that the diffraction of light is significant if the size of the void is comparable to the wavelength of the light.[22] Wavelengths of our interest lie in the range of 300-800 nm. Accordingly, we expect significant diffraction for wavelengths of interest if the size of a void ≃ 300 nm. Also, if the void size is less than the wavelength of the light incident, then the void acts as the circular source of the light.[22] Therefore, a) for unit cells with large width ($w = 10$ μm in fig 2b), the voids are also large in size and hence net absorption of photons in perovskite layer varies linearly with s due to the insignificant diffraction of light, b) for cells with small width ($w = 0.1$ μm in fig 2b), the absorption rate varies nonlinearly with s as the voids are also small in size (here the light incident on the void is distributed equally in all directions in the range -90° to 90° w.r.t. direction of incidence), and c) for cells with moderate width (i.e., $w = 1$ μm) the absorption rate varies linearly with $s$ for large voids (i.e., for $s < 60\%$) and non-linearly with $s$ for small voids (i.e., for $s > 70\%$). Interestingly, the optical simulations also show that the integrated photon absorption count in the perovskite for devices with '$s$' more than 70% along with unit cell size less than 1μm is similar to device with $s = 100\%$. This indicates that for devices with such small pinholes, the photon absorption is very efficient as there is significant diffraction of light. Hence, one may expect near ideal $J_{SC}$ for such cases. An analytical model to predict the effect of $s$ on the JV characteristics is discussed in the next section.

## 3.2 Electrical characteristics:

*Dark Current*: Since over the barrier transport of either type of carrier is negligible due to the presence of huge barrier potential in PSCs,[29] the dark current in the device is governed mainly by the recombination of charge carriers in the bulk of the perovskite or at the interfaces of different layers. Absence of any significant electric field fringing around the voids results in linear combination of dark current from ETL/HTL interface and bulk perovskite. Accordingly, the dark current density ($J_D$) can be written as

$$J_D(s) = sJ_{D_{EPH}} + (1-s)J_{D_{EH}}, \qquad (2)$$

where *EPH* in the subscript indicate presence of perovskite between ETL and HTL and *EH* indicate the absence of perovskite. Device with $s = 100\%$ forms p-i-n junction (ETL/Perovskite/HTL) and $s = 0\%$ forms p-n junction (ETL/HTL). SRH dominant recombination in the intrinsic layer of p-i-n structure and in the depletion region of p-n junction (interface traps) lead to ideality factor (*m*) close to 2[30] and therefore we do not expect much difference in the ideality factor in 100% surface coverage and 0% surface coverage device. Accordingly, eq. (2) can be expressed as

$$J_D = \left(sJ_{0_{EPH}} + (1-s)J_{0_{EH}}\right)\exp\left(\frac{qV}{mkT}\right) = J_0 \exp\left(\frac{qV}{mkT}\right). \qquad (3)$$

*JV characteristics under illumination*: The illuminated JV characteristics of any photovoltaic device can be written as sum of photocurrent ($J_{ph}$) and diode injection current ($J_{inj}$).[31] If principle of superposition holds then $J_{inj} = J_D$ and $J_{ph} = -J_{SC}$. It is shown in literature that proper doping of contact layers lead to superposition of light and dark JV characteristics,[26] therefore JV characteristics of the device under illumination ($J_l$) is given by

$$J_l = -J_{SC} + J_D. \qquad (4)$$

Using eqs. 3 and 4 with $J_l = 0$, we obtain the surface coverage dependent open circuit potential as

$$V_{OC}(s) = V_{OC_{EPH}} - \frac{mkT}{q}\ln\left[\frac{(1-s)\beta+s}{s}\right] + \frac{mkT}{q}\ln\left(\frac{\frac{J_{SC}(s)}{J_{SC_{EPH}}}}{s}\right), \tag{5}$$

where $\beta = \frac{J_{0_{EH}}}{J_{0_{EPH}}}$, $V_{OC_{EPH}}$ is open circuit potential and $J_{SC_{EPH}}$ is the short circuit current density of the device with $s = 100\%$ ($V_{OC_{EPH}} = \frac{mkT}{q}\ln\left(\frac{J_{SC_{EPH}}}{J_{0_{EPH}}}\right)$). Assuming IQE=100%[32], the short circuit current density is given by

$$J_{SC}(s) = q\frac{G(s)}{A}, \tag{6}$$

where $G(s)$ is the volume integrated generation rate (units: s$^{-1}$) and $A$ is the cross-sectional area. Accordingly, the lower limit of $J_{SC}(s)$ is,

$$J_{SC}(s) = sJ_{SC_{EPH}}. \tag{7}$$

Substituting for $J_{SC}(s)$ in eq. 5, we have

$$V_{OC}(s) = V_{OC_{EPH}} - \frac{mkT}{q}\ln\left[\frac{(1-s)\beta+s}{s}\right]. \tag{8}$$

Eq. 8 provides the lower limit of $V_{OC}(s)$. Comparison of equations 5-8 reveals that there could be significant difference in the $J_{SC}$ for similar $s$ but different void size distribution, but the difference in the $V_{OC}$ is very small (last term in equation 5).

*Model Predictions*: Though decrease in performance parameters of the device are expected with poor surface coverage, we observe some interesting features based on the analytical model as listed below-

a) **Effect on $V_{oc}$**: In general, reduction in $V_{OC}$ is expected in the devices with pinholes due to increased recombination in the device caused by ETL/HTL interface (2$^{nd}$ term in eqs. 5 and 8). Also eq. 5 indicates while the $V_{OC}$ has a strong dependence on surface coverage, the same has a rather weak dependence on the void size. For example, assuming β=10 along with m=2 result in ~ 100 mV change in Voc by changing s from 100% to 50% ($w = 1$ μm). But ±1 order change in $w$ (void size also changes by 1 order) for s=50% changes $V_{OC}$ only by ~ ∓10 mV.

b) **Effect on $J_{sc}$**: For large voids, $J_{SC}$ decreases with the decrease in surface coverage due to loss in absorption of photons caused by partial absence of active layer. But for devices with small voids, due to the increased generation rate in active layer (see fig 2b), we expect a nonlinear relation between $J_{SC}$ and $s$. Also, according to discussion related to fig 2b, we expect linear relation between $J_{SC}$ and $s$ for void size larger than ~400 nm.

c) **FF variation**: A detailed calculation for effect of $s$ on the FF is provided in supplementary material. According to our calculations and also from the general trend of $FF$ vs. $V_{OC}$, as reported in literature[33], we expect a decrease in FF with decrease in surface coverage.

d) **Efficiency**: Finally, we expect decrease in efficiency with decrease in surface coverage due to reduction in $J_{SC}$, $V_{OC}$, and $FF$. However, under limiting conditions (i.e., for devices with small voids or $\frac{J_{SC}(s)}{J_{SC_{EPH}}} \simeq 1$, and perfect interface between ETL and HTL) the third term in RHS of eq. 5 could offset the $V_{OC}$ loss due to pinholes (which is the second term of eq. 5). This indicates that nano-structured perovskite active layer could indeed result in better efficiency than planar devices with 100% surface coverage. .

*Simulation Results*: In order to validate our analytical model, we solved continuity and Poisson's equation using device simulation tool, Sentaurus.[34] Figure 3 shows the variation of $J_0$ and ideality factor with surface coverage as obtained from simulation. The direct interface between ETL and HTL is treated as recombination layer. The parameters used for simulation are provided in table S1. With these parameters our simulation results indicate that $J_0$ is higher for low surface coverage devices (see fig. 3a), which is an expected result from our analysis. We also observe linearity in $J_0$ with surface coverage and values predicted by analytical model (eq. 3 and solid line in fig 3a) are very close to simulation results. From our analysis, we expect that the ideality factor is independent of surface coverage and the same is observed from simulations (see fig 3b). As expected we also find from our simulations that dark JV characteristics are independent of void size distribution (see the discussion on the effect of unit cell width in Section II)

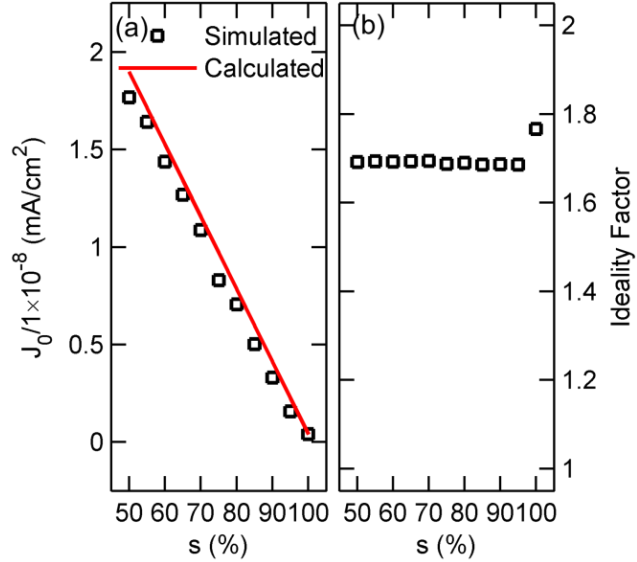

Fig 3: Surface coverage dependent dark JV characteristics. (a) $J_0$ and (b) ideality factor variation with $s$ as obtained from simulation. Variation of $J_0$ with $s$ according to analytical model (eq. 3) is also plotted in (a). For these calculations we have used 2 nm thick interface layer between ETL/HTL with corresponding SRH lifetime being 0.5 μs, while perovskite SRH lifetime is 2.73 μs.

Variation in $J_{SC}$ and $V_{OC}$ with surface coverage, obtained from JV simulation under illumination, are shown in figure 4. Part (a) and (b) shows the variation in $J_{SC}$ and $V_{OC}$ with surface coverage ($w = 1$ μm, $\beta = 94$). Here we assume that the photo-generation rate varies linearly with the surface coverage (see the lower limit calculation in Fig. 2). We find that the $J_{SC}$ varies linearly with surface coverage, which indicates that the charge collection efficiency under short circuit conditions is not influenced by surface coverage. The corresponding open circuit voltage is plotted in figure 4b. We observe a close agreement between simulation results and analytical expressions. For $V_{OC}$ calculations we use ideality factor 1.7 as obtained from dark JV simulations. A sharp increase in $V_{OC}$ for higher surface coverage (near 100%) indicates that the presence of very small fraction of pinholes or voids can affects $V_{OC}$ to a large extent. A decrease in FF with increase in $J_0$, due to presence of bad ETL/HTL interface, in low surface coverage devices is expected and the same has been confirmed from the simulation results also (see fig. S4 for details).

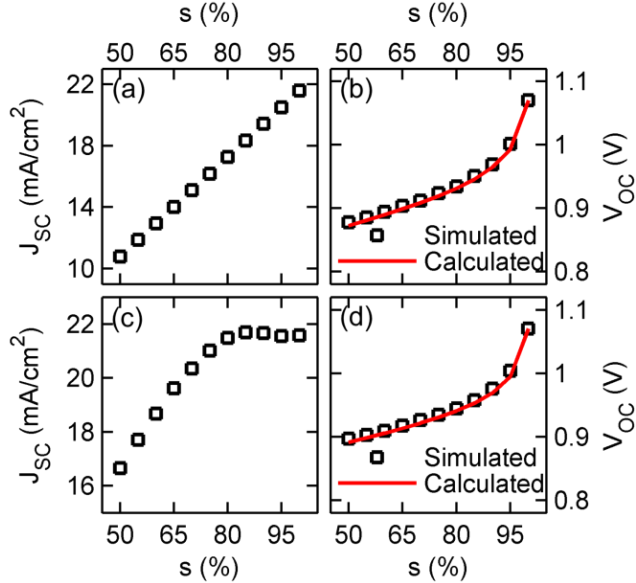

Fig 4: Effect of surface coverage on the light JV characteristics. (a) and (b) show the effect of $s$ on $J_{SC}$ and $V_{OC}$ when void size is large enough such that diffraction of light does not have significant effect on the generation rate in the active layer. These results serve the purpose of lower limit of $J_{SC}$ and $V_{OC}$ with $s$. (c) and (d) show the effect of $s$ on $J_{SC}$ and $V_{OC}$ when void size is small and there is significant diffraction of light ($w = 1$ μm). In this case, we observe saturation in current density near 80% surface coverage. We find that in both cases simulated results (squares) are in close agreement with the results obtained from analytical calculations (solid line).

The optical simulation results in Fig. 2 indicated that the photon absorption rate and hence the carrier generation rate is indeed dependent on the surface coverage and size of voids. Accordingly, we performed the device electrical simulations for $w = 1$ μm with carrier generation rates obtained from the corresponding optical simulations (see fig 2b). The variation of $J_{SC}$ and $V_{OC}$ with the surface coverage for $w = 1$ μm are plotted in fig 4c and 4d. We find significant changes in $J_{SC}$ values as compared to the limiting case (see Fig. 4a) while changes in $V_{OC}$ values are nominal. We also observe that $J_{SC}$ varies nonlinearly for small voids (i.e., for $s > 70\%$) and linearly with large voids (i.e., for $s < 65\%$) – which are similar to the trends observed for the integrated photon absorption rates (compare fig 2b and 4c). This observation supports our assumption of IQE=100% for generalized case (see eq. 6). Further, we find that calculated values of $V_{OC}$ using equation 5 are in close agreement with the values obtained from simulation (see fig 4d). According to our analysis the difference in limiting $V_{OC}$ and actual $V_{OC}$ for w=1 μm at s=50% is 18.8 mV which is in close agreement with the simulated value (19 mV). Our simulation results also confirm the weak dependency of $V_{OC}$ on void size distribution as compared to surface coverage. The corresponding FF and efficiency are plotted in fig S4 and we observe increasing trends for both with $s$.

### 3.3 Effect of HTL/ETL interface recombination:

We observe from eqs. 5 and 8 that $V_{OC}$ of the device decreases with increase in $\beta$. Also, $\beta$ is inversely proportional to the carrier lifetime at the interface of ETL/HTL ($\tau_{interface}$) when all other parameters are kept constant (because $J_0$ is inversely proportional to carrier lifetime[29]). Equation 5 indicates that 1 order of magnitude increase in $\tau_{interface}$ results in 100 mV increase in open circuit voltage when $\tau_{interface} < \tau_{bulk}$ ($s = 80\%$), while $\tau_{interface} > \tau_{bulk}$ results in saturation of $V_{OC}$ to $V_{OC_{EPH}} + \frac{mkT}{q} \ln\left(\frac{J_{SC}(s)}{J_{SC_{EPH}}}\right)$ –a value more than the open circuit voltage of device with $s = 100\%$ ($V_{OC_{EPH}}$). This is because $\beta(1 - s) + s \rightarrow s$ in this regime (eq. 5). Our simulations also show the same trends and representative data for $V_{OC}$ and efficiency with interface lifetime for 80% surface coverage ($w = 1$ μm) are plotted in figure 5a. We observe logarithmic increase in $V_{OC}$ with the ETL/HTL interface lifetime. Also we find 7 mV improvement in $V_{OC}$ as compared to device with $s = 100\%$ for $log(\tau_{interface}/\tau_{bulk}) = 2.3$ while the limiting value, in the absence of ETL/HTL recombination ($\beta = 0$), is ~9 mV. We observe a rather small improvement in efficiency as well for this case. These results indicate that nanostructured perovskite solar cells, with appropriate interface engineering, could be an interesting prospect and are in broad agreement with the higher efficiency for low surface coverage devices reported in literature.[15] Curiously, we find that interface recombination at perovskite/HTL interface do not affect the performance in any significant manner. A representative data showing the variation of $V_{OC}$ with s for 0.1 μs effective lifetime for perovskite/HTL interface is plotted in fig 5b.

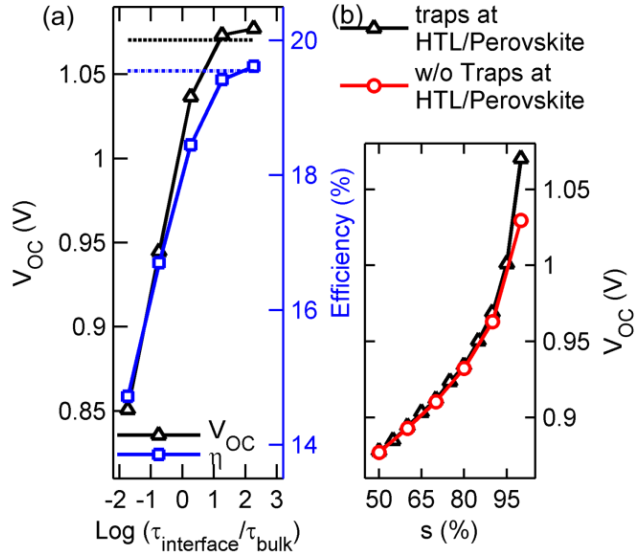

Fig 5: Effect of interface lifetime on the performance parameters of the device. (a) The effect of ETL/HTL interface lifetime ($\tau_{interface}$) on $V_{OC}$ and efficiency of the device with 80% surface coverage ($w = 1$ μm). Here $\tau_{bulk}$ is the SRH lifetime of the perovskite (2.73 μs). We observe a slight increase in $V_{OC}$ for very high velue of ETL/HTL interface SRH lifetime. (b) The effect of perovskite/HTL interface lifetime on $V_{OC}$ of the device as a function of surface coverage ($s$). We observe an insignificant change in $V_{OC}$ of the device in the presence of traps at perovskite/HTL interface.

**3.4 Characterization technique to identify surface coverage**:

Finally, we provide a technique to identify the surface coverage of a PSC. For this purpose we rewrite the eq. 5 as,

$$s = \frac{\beta - J_{SC_r} \exp(v_{OC})}{\beta - 1}, \tag{9}$$

where $J_{SC_r} = \frac{J_{SC}(s)}{J_{SC_{EPH}}}$ and $v_{OC} = \frac{V_{OC_{EPH}} - V_{OC}(s)}{mkT/q}$. The parameters $\beta, J_{SC_{EPH}}, V_{OC_{EPH}}$, and $m$ can be estimated independently by fabricating devices with known surface coverage (See supplementary material for details). These parameters along with $J_{SC}(s)$ and $V_{OC}(s)$ can be used to obtain very accurate estimates for s using eq. 9. For example, if $\beta = 100, J_{SC_{EPH}} = 22$ mA/cm², $V_{OC_{EPH}} = 1.1$ V, and $m = 2$ for a given batch of devices along with $J_{SC}(s) = 18$ mA/cm² and $V_{OC}(s) = 0.9$ V then eq. 9 indicates that the surface coverage could be 60%. Note that the upper limit of s is $J_{SC_r}$ (see discussion on lower limit of $J_{SC}(s)$ and eq. 7). Hence, any larger value for $s$, as predicted by eq. 9, indicates inaccuracies related to the basic parameter extraction (*i.e.*, in $\beta, J_{SC_{EPH}}, V_{OC_{EPH}}$, and $m$).

## 4. Conclusions

To summarize, here we explained the dependency of perovskite solar cell performance on its surface coverage through detailed optoelectronic modelling. We identified that – a) loss in $J_{SC}$ is directly proportional to the loss in generation rate which is affected by void size distribution as well as the surface coverage, b) unlike $J_{SC}$, $V_{OC}$ is strongly dependent on surface coverage rather the void size, and c) FF decreases with the decrease in surface coverage. Further, our simulations indicate that in limiting case and with proper interface engineering (*i.e.* with zero or negligible ETL/HTL interface recombination), it is indeed possible to have a device with suboptimal surface coverage but better efficiency as compared to 100% surface coverage device. Finally, we have presented a simple technique to characterize the surface coverage in perovskite solar cells –a result that could be of immense interest to the community.

## Acknowledgement

This paper is based upon work supported under the US-India Partnership to Advance Clean Energy-Research (PACE-R) for the Solar Energy Research Institute for India and the United States (SERIIUS), funded jointly by the U.S. Department of Energy (Office of Science, Office of Basic Energy Sciences, and Energy Efficiency and Renewable Energy, Solar Energy Technology Program, under Subcontract DE-AC36-08GO28308 to the National Renewable Energy Laboratory, Golden, Colorado) and the Government of India, through the Department of Science and Technology under Subcontract IUSSTF/JCERDC-SERIIUS/2012 dated 22$^{nd}$ November 2012. The authors also acknowledge the Center of Excellence in Nanoelectronics (CEN) and National Center for Photovoltaic Research and Education (NCPRE), IIT Bombay, for computational facilities.

Supplementary material for "Pinhole induced efficiency variation in perovskite solar cells" can be provided upon request